\documentclass[final]{revtex4}
\usepackage{graphicx}
\usepackage{amssymb}
\usepackage{bm}

\begin{document}

\title{Immersive VR Visualizations by VFIVE. Part 1: Development}

\author{Akira Kageyama}
\address{Graduate School of System Informatics, Kobe University, Kobe, 657-8501, Japan\\
kage@cs-kobe-u.ac.jp}

\author{Nobuaki Ohno}
\address{University of Hyogo, Kobe, 650-0047, Japan}

\begin{abstract}
We have been developing a visualization application for CAVE-type 
virtual reality (VR) systems for more than a decade.
This application, VFIVE, is currently used in 
several CAVE systems in Japan
for routine visualizations.
It is also used as a base system of further developments of
advanced visualizations.
The development of VFIVE is summarized.
\end{abstract}


\maketitle

\section{Introduction\label{sec:intro}}

CAVE is a room-sized, immersive virtual reality system
developed in early 1990s at Univ.~Illinois, Chicago\cite{Cruz-neira1993}.
Wall screens on which stereoscopic images are projected from the rear 
surround the viewer in the room.
The floor is another screen on which stereo images are projected
usually from the ceiling.
A high-precision head tracking system with the six degrees-of-freedom (DOF) is installed.
The tracker's data are sent to a graphics system
to generate natural scenes for the viewer.
A portable controller, sometimes called wand, is used 
for the human machine interface in the CAVE's room.
The wand is also tracked.
Compared with other virtual reality (VR) systems,
such as the head-mounted display system,
CAVE provides much wider view and higher immersive sense.
The scientific visualization is
one of the most important application fields
from the first CAVE\cite{Cruz-neira1993}
to the latest generation called StarCAVE\cite{Defanti2009a}.

One of the early visualization tools for CAVE systems
was CAVEvis\cite{Jaswal1997} 
developed in 1990s, which was
to analyze time varying fluid data.
Since then, CAVE systems have been used in broad spectrum of visualizations,
including seismic simulation\cite{Chopra2002},
meteorological simulation\cite{Ziegeler2001},
biomedical fluid simulation\cite{Forsberg2000},
magnetic resonance imaging\cite{Zhang2001},
turbulence simulation\cite{Tufo1999},
CFD of molten iron\cite{Fu2010},
CFD of wind turbines\cite{Yan2011},
geomagnetic field\cite{Bidasaria2005},
and archaeological\cite{Acevedo2001}.
Various applications in scientific visualization and related
user-interface study at Brown University are summarized
in\cite{Jr2009}.
An application of the StarCAVE\cite{Defanti2009a} in geophysics can be 
found in\cite{Lin2011}.

We also started developing visualization tools for 
a CAVE system in the late 1990s for visualization 
of our simulations on plasma physics and related fields\cite{kageyama1998}.
Our visualizations in the CAVE gradually attracted interests 
of other simulation researchers.
By their request, we developed several CAVE programs
for different kinds of simulation data.
Through the experience of those developments,
we had found that several methods and tools could be used in common.
We combined them into an general-purpose visualization
application for field-type data\cite{kageyama1999,Kageyama2000}.
Since then, we have been improving VFIVE
for more than a decade. 
The continued improvements
have made this application being a matured, practical tool 
for three-dimensional visualization in CAVE systems.

To the authors' knowledge, VFIVE is unique 
visualization application for CAVE systems 
with regard to its long continued development history.
Another characteristic feature of VFIVE is that
it has always been developed in a ``request-driven'' way.
We have added new functions or visualization methods
only when they are strongly required by simulation researchers.
This policy has played an important role
to keep VFIVE being a simple, minimum, but still practical tool.

The purpose of this paper is to summarize the development
of VFIVE.
Implementations and usages of various visualization 
methods in VFIVE are outlined.

VFIVE is coded with C++ language with standard APIs of OpenGL and CAVElib.
The source codes are freely distributed.
It runs both on Linux-based workstations
and Windows-based PC clusters.
VFIVE is now used in various CAVE systems in Japan.
It is also used as a base code set for further 
developments of advanced visualizations.
Those applications and 
scientific achievements made possible by VFIVE 
in several CAVE systems are described in 
our another paper\cite{Kageyama}.

%
\section{VFIVE's User Interface}
%

\begin{figure}[h]
  \begin{center}
   \includegraphics[width=0.30\linewidth]{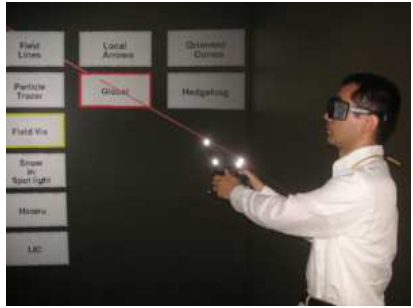}
  \caption{%
    Menu function of VFIVE. 
    The menu panels appear in the CAVE's VR space. 
    The user shoots one of the menu panels by a virtual laser beam 
    emitted from the wand controller 
    to select the data and visualization method.  
  }
  \label{fig:v5menu}
  \end{center}
\end{figure}

We have implemented a menu-type user interface in VFIVE.
Each visualization method implemented in this application
is selected by shooting a virtual menu panel with a beam from the wand.
When a user in the CAVE room
presses a button of the wand,
a virtual laser beam appears from the tip of the wand.
At the same time,  menu panels appear in front of the user (see Fig.\ref{fig:v5menu}). 
By shooting submenu panels one after another, a target data to be visualized, 
then a visualization method to be applied to it are selected.
Visualization parameters corresponding to the selected visualization
method are specified by the wand's another button.
The parameters and the interface in each visualization methods 
are described in section~\ref{subsec:vm}.

\begin{figure}[h]
 \begin{center}
   \includegraphics[width=0.3\linewidth]{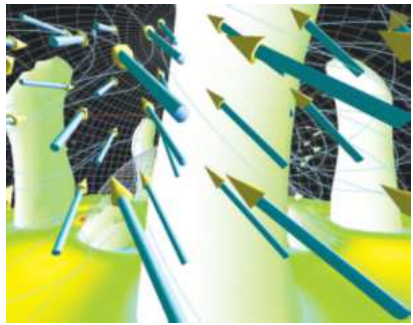}
  \caption{%
    VFIVE's visualization method Isosurface (white objects),
    with Local Arrows,
    Field Lines (thin light-blue curves),
    and Orthoslicer (horizontal plane).
   }
  \label{fig:isosurface}
 \end{center}
\end{figure}

\begin{figure}[h]
 \begin{center}
   \includegraphics[width=0.5\linewidth]{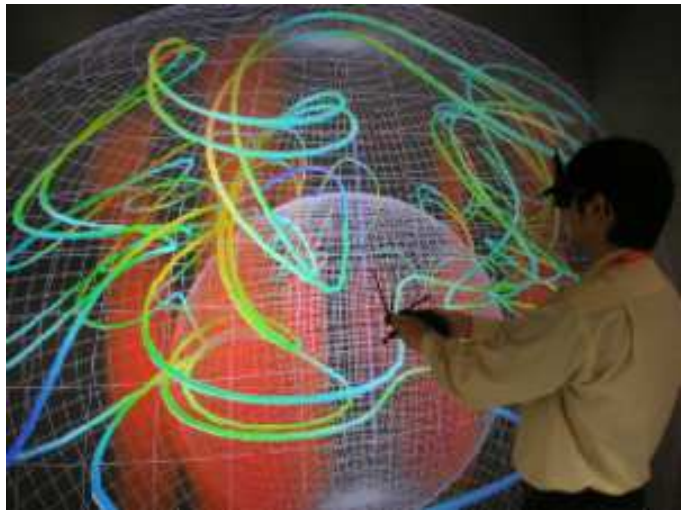}
  \caption{%
    Volume rendering of a scalar (axial vorticity)
    in the geodynamo simulation. The three-dimensional
    texture mapping technique is used.
    Tubed streamed lines are also shown,
    which visualize the convection velocity.
  }
  \label{fig:tube_and_volren}
 \end{center}
\end{figure}

Multiple visualization methods thus selected can be superimposed to show
a combined view as shown in Fig.~\ref{fig:isosurface}.
In this case, Isosurface, Orthoslice, Local Arrows, and 
Tracer Particle methods are shown. 
These methods will be described later.
Fig.~\ref{fig:tube_and_volren} is a snapshot of geodynamo simulation
visualization in which tube-shaped Field Lines of the
flow and Volume Rendering of a scalar field are superimposed.

%
\section{Visualization Methods in VFIVE\label{subsec:vm}}
%
\subsection{Particle Tracer}
\begin{figure}[t]
  \begin{center}
   \includegraphics[width=0.5\textwidth]{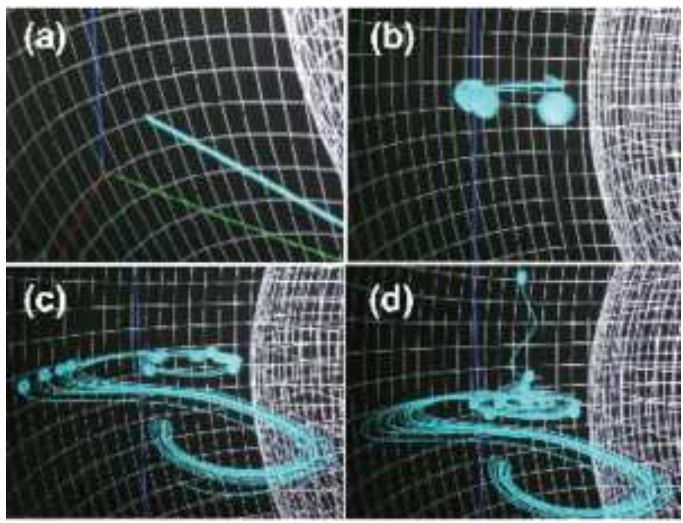}
  \end{center}
  \caption{%
    A visualization method Particle Tracer of VFIVE for the CAVE VR System. 
    A sequence of seed points of streamlines are placed 
    by repeated clicking of the wand's button.
    Convection flow in the Earth's liquid core is visualized
}
  \label{fig:tracer3}
\end{figure}

After shooting a virtual menu panel
saying ``Particle Tracer'' and pressing the wand button,
a short blue beam appears from the wand.
The tip of the blue beam designates the seeding point of 
a streamline of the vector field that is already selected by the menu.
The tracer particle calculation is 
performed with a 6-th order Runge-Kutta integration method in
real time; 
the vector value at the tracer particle position 
is interpolated in real time and 
the user can see the particle is moving (or flying) 
in the CAVE's VR space.
To show the streamline, 
a curve is drawn in the trail of the particle.
In this method, the speed of the particle indicates the vector field strength
at the position.
Fig.~\ref{fig:tracer3} shows a sequence of snapshots of the Particle Tracer visualization
of a geodynamo simulation performed on the Earth Simulator supercomputer

A useful way to apply this method is to 
place a set of seed points
in a small local region of the target vector field
by a rapid, successive clicking of the wand's button, 
slightly shifting the beam tip.
Then, a bunch of tracer particles are born and fly like a flock of birds.
They move as a group for a while 
if the initial target area is small enough,
but the distances between the particles exponentially grow
when the target vector field is chaotic that 
is common in nonlinear fluid simulations as our goeynamo simulation.
We sometimes notice that some of the tracer particles show attractive
behavior such as a sharp turn, helical winding, or other dynamical and intriguing twists.
The place where the tracer particles show such characteristic
changes is usually a ``hot spot'', i.e., a key area of the target vector 
field in order to uncover the underlying physics.
Then we walk in the CAVE room or make a virtual fly in the space by the 
wand's joystick to get closer to the hot spot.
One may apply the Particle Tracer analysis again around the hot spot
to analyze the structure of the vector field in detail 
in and around the hot spot.
Other visualization methods described below could be applied, too.
Our experiences have told that 
the Particle Tracer is an effective visualization method
to find a hot spot of the vector field and to intuitively understand
the three-dimensional structure of the vector field.

\subsection{Field Line}
This visualization method is almost the same one as the Particle Tracer described above.
The differences are; 
(i) the 6-th order Runge-Kutta integration is
performed in both positive and negative directions,
and (ii) the target vector field is normalized before the integration
so that the vector amplitude is unity everywhere in the space.
We discard unnecessary information of the vector amplitude since
we are interested only in the distribution of 
the direction (angles) of the vector field in this 
visualization method.

\subsection{Local Arrows}
This is another visualization method for vector fields.
After selecting Local Arrows from the menu and pressing a wand button,
tens of arrows appear around the tip of a short (virtual) beam,
within a spherical region.
The size and angle of each arrow indicate
the magnitude and direction of the vector field.

%
\begin{figure}[h]
  \begin{center}
   \includegraphics[width=0.5\textwidth]{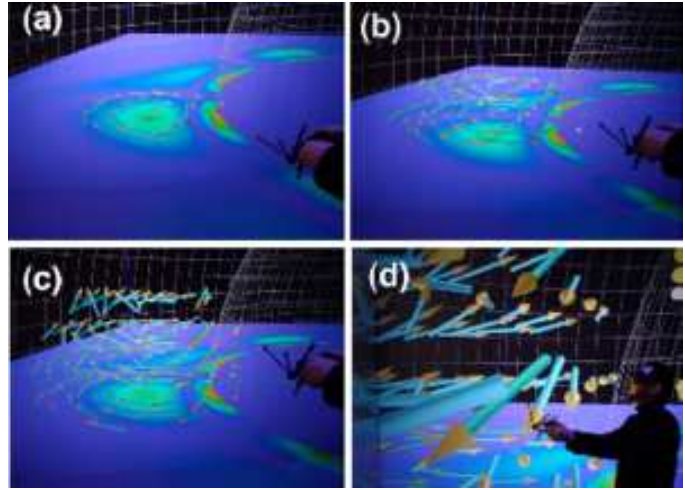}
  \caption{%
    VFIVE's visualization method Local Arrows (3-D arrow objects with blue and yellow).
    A group of arrows follows the wand motion. 
    Here, the user is moving up the wand from (a) to (c). 
    The arrows are observed in detail by a closer look in (d). 
    The interpolations of vector field for each arrow are performed in real time. 
    Orthoslicer (horizontal plane) is also shown in this figure.
 }
  \label{fig:arrows3}
  \end{center}
\end{figure}
%

This is an example of fully interactive or dynamic visualization method
that makes use of the CAVE's interactivity.
As the user moves the wand, 
the bunch of arrows follows the hand motion.
See Fig.~\ref{fig:arrows3}.
The real time spatial interpolation from the target grid data is taken.
The user can intuitively understand, or ``feel'' the 
local distribution of the vector field 
through the combination of the hand motion and 
three-dimensional observation of the dynamical changes 
like ``dancing'' of the arrows.

\subsection{Snowflakes}
\begin{figure}[h]
  \begin{center}
   \includegraphics[width=0.75\textwidth]{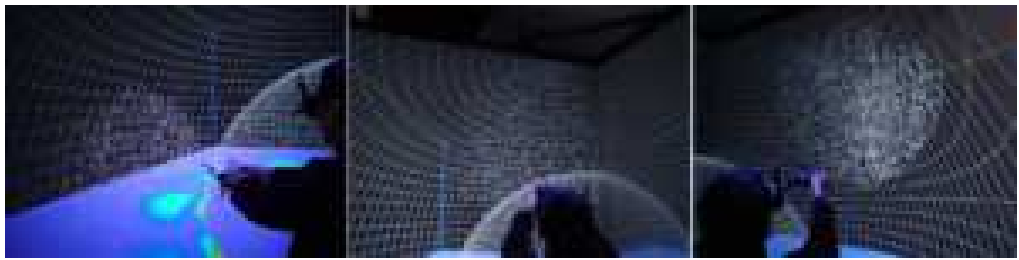}
  \end{center}
  \caption{%
    VFIVE's visualization method ``Snowflakes''.
    Hundreds of tracer particles in a cone-shaped spotlight emitted from the wand. 
   }
  \label{fig:snows}
\end{figure}

Visualization method called Snowflakes (Fig.~\ref{fig:snows}) 
shows hundreds of tracer particles in a cone-shaped spotlight.
The apex of the cone is the wand and the axis is 
always parallel to the wand direction.
The cone-shaped spotlight follows the wand's motion.

%
%

This visualization method is inspired by, and named after, 
a scene in a snowing night observed by one of the authors.
He noticed that snowflakes flying under a streetlight
effectively visualize the wind there.
This method is used to dynamically changing the wand 
direction---like using a searchlight---and observing 
the particles in the spotlight.

\subsection{Isosurface}

Isosurface is one of standard visualization methods for
scalar fields.
We use the classical marching cubes algorithm\cite{Lorensen1987}
for the isosurface construction in VFIVE.
The white objects in Fig.~\ref{fig:isosurface} are 
isosurfaces of a scalar field 
in the geodynamo simulation.

To control the isosurface level or threshold,
a three-dimensional virtual slider object appears near the user
(not shown in Fig.~\ref{fig:isosurface})
when a wand button is pressed.
The virtual slider moves vertically 
in accord with the wand's vertical motion.
It changes isosurface level.
When the button is released, the threshold is determined 
and the marching cubes calculation begins before
new isosurface object is shown.

\subsection{Orthoslicer}
Orthoslicer is another popular visualization methods for scalar fields.
After selecting the Orthoslicer panel from the menu,
a box-shaped small object, which symbolizes the whole simulation data space, 
appears in the CAVE room when a wand button is pressed. 
A mesh in the symbol box designates the slice position.
To control the slice plane,
all one has to do is to move the wand in the desired direction,
with pressing a wand button.
The mesh in the box moves in accord with the wand motion.
When the wand's button is released,
the distribution of the scalar data is rendered at the specified slice
position in the data.
Orthoslicers can be seen with other visualization 
methods in Fig.~\ref{fig:arrows3} and \ref{fig:isosurface}.

\subsection{Volume Rendering}

One of technical challenges in the CAVE visualization
is implementation of the volume rendering
which is a popular as well as very powerful 
visualization method for scalar field\cite{Drebin1988}.
The volume rendering is 
a heavy and demanding numerical process, since
it requires numerical integrations on each pixel when
the ray casting, that is classical algorithm in this method, is adopted.

To realize a high-speed volume rendering in VFIVE,
we use the three-dimensional
texture mapping technique\cite{Cullip1993}.
In this method, 
polygon slices draped with semi-transparent textures derived from 
the target scalar field are piled 
up from back to front. 
A sample snapshot of the CAVE volume rendering with this method
in VFIVE is shown in Fig.~\ref{fig:tube_and_volren}.

In volume renderings,
the color and opacity is determined from
the scalar field through the so-called
transfer functions.
In our implementation the transfer function
can be edited in the CAVE room interactively with a 3-D GUI made for this purpose.
An interaction panel appears in the CAVE's VR space
to control the parameters of the transfer function through wand motion.
Details of the implementation and its GUI in VFIVE are
described in\cite{Ohno2007}.

\section{Animation}
%

VFIVE was originally designed to analyze stationary fields, without the time variation.
When the target simulation is non-stationary,
we have used VFIVE to analyze a snapshot data 
or three-dimensional field at one specific time.
To resolve this limit, 
we have implemented an animation function in VFIVE.
Once ``Animation" is selected in the menu, VFIVE reads a 
four-dimensional data from the hard disk drive (HDD).
The data consists of a time sequence of three-dimensional data,
from the first time step at, say $t=1$, to the final time step at $t=N$.
In the beginning, the VFIVE shows the data at the first time step $t=1$
in the CAVE room.
The user chooses and applies arbitrary visualization methods
described above (and below)
by making use of the menu and the wand to specify
the visualization method and parameters.
The visualized results are shown in the CAVE as usual,
but at the same time, the visualization objects
(polygons) are saved in the HDD.
Then the same visualization methods are applied to
the data of next time step at $t=2$,
with the same parameters.
The visualization results are shown in the CAVE,
and again the polygonal data for $t=2$ are saved in the HDD.
The same procedure is applied until $t=N$.
After generating all of the polygonal data from $t=1$ to $N$,
VFIVE begins displaying the sequential polygonal data
one after another, reading the data from the HDD.

%
\section{VTK Combination}
%

\begin{figure}[htbp]
  \begin{center}
   \includegraphics[width=0.5\linewidth]{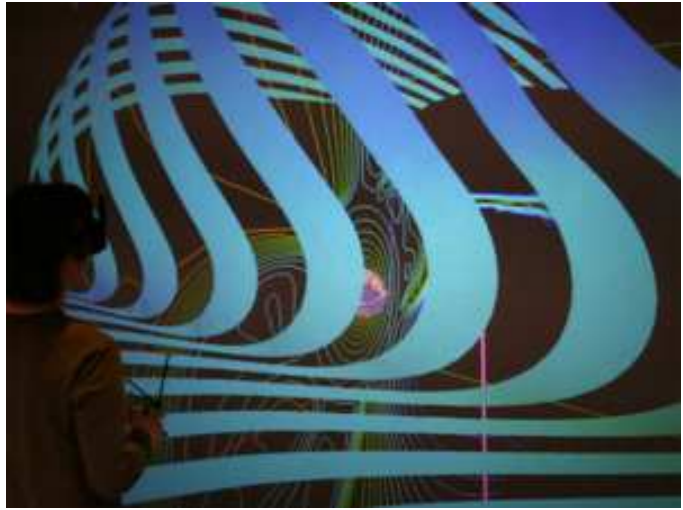}
  \caption{%
    VTK's Stream Surface and Contour Lines integrated in VFIVE.
    Flow velocity and pressure of a magnetosphere simulation are visualized.
  }
  \label{fig:magneto_ribbon}
  \end{center}
\end{figure}

``VTK" (Visualization Tool Kit)\cite{Schroeder2004}
is a rich and sophisticated visualization library.
One can draw out the polygonal data of
visualization objects of VTK, which can be independently rendered by OpenGL.
Applying this technique, we have
incorporated a lot of (polygon-based) visualization methods such as Tubes and Contour Lines of
VTK into VFIVE\cite{Ohno2006}.
An example of VTK visualization is shown 
in Fig~\ref{fig:magneto_ribbon},
which is a visualization of a magnetohydrodynamic
simulation of the Earth's magnetosphere in the CAVE.
In this simulation, time development of multiple vector fields
(such as velocity field and magnetic field)
and multiple scalar fields (such as pressure and density)
are numerically solved.
The velocity field is visualized by ``Stream Surface" 
and the pressure is visualized by contour lines.

We have developed adequate GUIs for every VTK method that is 
incorporated into VFIVE,
in accordance with other interface of VFIVE.
For example, GUI of VTK's contour lines are designed
as the same one as that of Orthoslicer described above.
GUI for VTK Tubes is the same as VFIVE's Particle Tracer and Field Lines. 
For the VTK's Stream Surface,
the wand and a beam are used to specify two points that are necessary for
this visualization method.
By pressing the wand's middle button, the first seed is placed 
at the tip of beam and the second seed is placed when the button is released.

We have found that 
users of VFIVE tend to prefer 
VFIVE visualization methods than VTK-original methods.
This would be because the visualized 
objects produced by VTK are basically ``still'' one,
in contrast that VFIVE's other methods generate dynamically
changing objects such as flying Particle Tracers or 
hand-following Local Arrows.
From this reason, the VTK-enabled version of VFIVE has branched off
from the major VFIVE development trunk now.

%
\section{Region of Interest and Level of Detail
\label{sec:roi}}
%

When we use a CAVE for the visualization of large-scale data,
there are two different kinds of purposes.
One is to grasp the whole structure of the three-dimensional field
of the input data to be analyzed,
and another is to understand the detailed structure 
as accurate as possible at a small but 
important spatial region, or a hot spot.
In order to make these complemented
purposes be compatible,
we have implemented
the function of interactive region of interest (ROI) 
linked with level of detail (LOD) in VFIVE.
Details of this implementation are described in\cite{Ohno2010}.

%
\section{Other Methods and Functions}
%

In this subsection, we briefly introduce
other functions and methods that are not mentioned so far.

Since our goal is to visualize large-scale simulation data
produced by supercomputers,
the data size is always our challenge.
To handle large-scale data within acceptable time,
we have parallelized VFIVE by OpenMP to accelerate calculations.

In Local Slicer,
the user holds and moves a small rectangle plane in the CAVE room.
The plane color shows the distribution of
a scalar field, which is selected by the menu, on the slice.
As in the Local Arrows, the Local Slicer follows the wand motion.
There are two modes of the Local Slicer;
in the first mode, the normal vector of the plane is always perpendicular
to wand direction, and in the
second mode it is always perpendicular to a specified 
vector field.
Near a ``hot spot'' of the field where
the vector and scalar fields change sharply,
one can observe a dynamic motion of the plane as
he/she slowly moves the wand around there.

In the visualization method named Hotaru, which means firefly in Japanese,
thousands of tracer particles are randomly scattered in the whole
simulation data space.
The 1st order Euler method is used for the numerical
integration since statistical behavior rather than 
the accuracy of each motion is important in this visualization.

\begin{figure}[htbp]
 \begin{center}
   \includegraphics[width=0.5\linewidth]{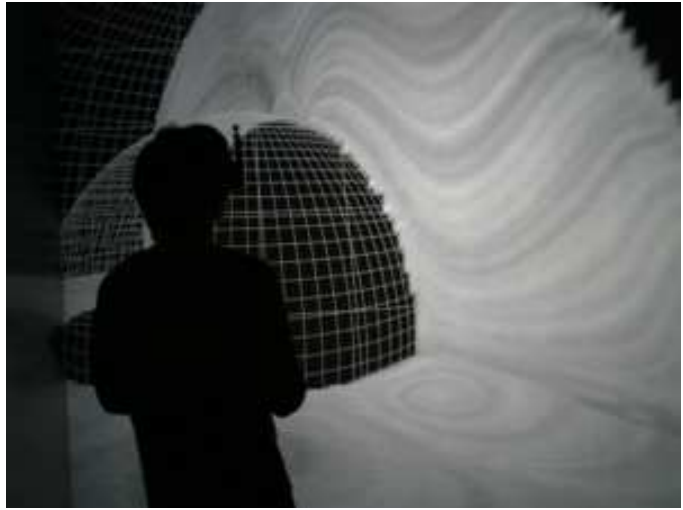}
  \caption{%
    Line Integral Convolution (LIC) method applied to the geodynamo simulation. 
    LIC is used to visualize the convection velocity.
  }
  \label{fig:dynamo_lic}
 \end{center}
\end{figure}

The Line Integral Convolution (LIC) is a useful visualization
method for vector fields.
We have incorporated this method into VFIVE.
The LIC method is applied to a specified vector field
by the menu with the wand at a slice plane 
whose position is also controlled by the wand.
The CAVE GUI for LIC method is the same as that for the Orthoslicer;
the user can interactively select the 
slice plane of the LIC by moving the wand in horizontal or
vertical directions.
Fig.~\ref{fig:dynamo_lic} shows an example of the LIC visualization 
applied to a CAVE visualization of the geodynamo simulation.

VFIVE has a minimum function to display three-dimensional objects,
composed by triangles and line segments. 
This function is used to display, for example, the boundary
planes of the simulation.

Snapshot is a function to save the frame buffer to the HDD.
Images displayed on the front wall screen are saved to 
sequential files when the user selects this option.
This is useful to store the image data visualized by VFIVE.
Various visualization parameters such as
isosurface threshold, seeding positions of Field Lines and Tracer Particles,
slice positions of Orthoslicer are also saved for later
usage by VFIVE or other visualization software on other 
platforms like PCs.

\begin{figure}[htbp]
 \begin{center}
   \includegraphics[width=0.5\linewidth]{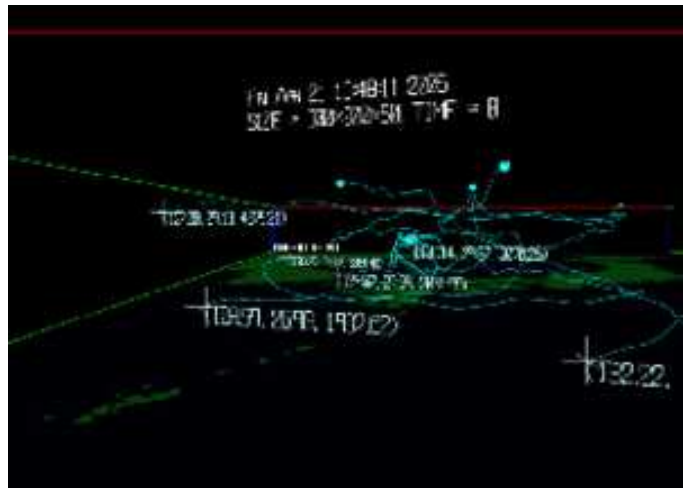}
  \caption{%
    Text strings in VFIVE.
    Visualization parameters such as seeding positions
    of each tracer particle and the time step of the animation
    are shown.
   }
  \label{fig:strings}
 \end{center}
\end{figure}

VFIVE can display texts in the CAVE room by our hand-made stroke font.
This function is used to display visualization parameters
by text strings.
In Fig.~\ref{fig:strings},
seeding positions of streamlines
are shown.
The time steps in the animation mode
and simulation data size are also displayed
in the CAVE's VR space.

%
\section{Conclusion\label{sec:conclusion}}
%
We have developed a visualization application, VFIVE, for CAVE-type VR systems
for immersive and interactive
visualization of large-scale simulation data.
VFIVE makes it possible to analyze multiple scalar and
vector fields at the same time in the CAVE.

In its development, special emphasis has been placed on the interactivity, since
we believe that the interactivity is the key
in the visualization\cite{Hibbard1989,Lum2002},
especially in the CAVE systems.
Take the particle tracer or field lines, for example, that are 
common and standard visualization methods
implemented in many visualization tools on PC-based software.
It is, however, rather difficult to accurately specify the seed positions 
by the standard mouse-based GUI.
What we want to do in the vector 
field visualization is to get the answer to a question like:
``what if we seed a tracer particle at the center of these two vortices?'',
rather than a question like:
``what if we seed a tracer particle at the position $(x,y,z)$?''
The CAVE with the head and hand tracking system enables us
to perform effective and accurate seeding in those situations.

In addition to the effective and accurate
positioning of the seed points,
the tracer particle or field line
in the CAVE is qualitatively different
from those in PC-based visualization software.
In the CAVE room, we can observe
tracer particles in motion 
and if a particle shows a sharp and intriguing change 
of the orbit, we can literally walk to that place in the CAVE room
and analyze the vector field
there in detail, while the particle is still
moving around there.

The source codes of VFIVE are freely available.
It is actively used at several CAVE systems in Japan 
for various kinds of visualizations.
Those applications and 
scientific achievements made possible by VFIVE are summarized in
another paper\cite{Kageyama}.

%
\section*{Acknowledgement}
%
This work was supported by
JSPS KAKENHI Grant Number 23340128
and Takahashi Industrial and Economic Research Foundation.

%

\end{document}